# Solar GES-structure modified with EiBI gravity

## Souvik Das and Pralay Kumar Karmakar[*]

*Department of Physics, Tezpur University, Napaam, Tezpur, Assam-784028, India*

[*]Corresponding author.

*E-mail address:* pkk@tezu.ernet.in, pkk170722@gmail.com

ABSTRACT

In the post-Newtonian era, the Eddington-inspired Born–Infeld (EiBI) theory, considered as an improved modification of the Einsteinian general relativity formalism in the weak field regime (non-relativistic), has enabled us to study the dynamics of dense astroobjects in light of the modified gravitational effects. This EiBI theory imparts a new shape to the usual gravitational Poisson equation through the addition of a cosmological correction factor, termed as the EiBI gravity parameter. A systematic inclusion of this gravity in the basic structure equation could lead to a realistic picture of the existing solar models free from any end-stage singularity. A theoretic model is accordingly proposed to investigate the effect of the EiBI gravity on the Gravito-Electrostatic Sheath (GES) formalism of the equilibrium solar plasma structure. This study shows that the GES-based solar plasma dynamics is noticeably modified against the previously reported Newtonian GES-model studies. An equilibrium bounded solution for the solar self-gravity shows the EiBI-modified solar surface boundary (SSB) to exist at a new helio-centric radial location $\xi = 4$ (on the Jeansean scale). It is found that the EiBI gravity shifts the present SSB outwards by 14.28% relative to the original Newtonian SSB. The EiBI-modified gravity effects on diverse relevant solar parameters, such as the gravito-electrostatic potentials, fields, and Mach numbers, are illustratively analyzed. It is anticipated that our analyses could be applied further to see the solar plasma equilibrium and fluctuation dynamics in realistically modified post-Newtonian gravity environments on both the bounded (interior) and unbounded (exterior) solar plasma scales.

## 1. Introduction

The Sun is a complex, dynamic, and self-gravity-confined naturalistic plasma system powered by thermonuclear fusion reactions in its core [1,2]. The solar plasma system, confined by the joint action of its strong intrinsic self-gravity and magnetic field, has been a fundamental issue to study diverse characteristics of the Sun and its circumvent atmosphere in the form of so-called solar wind for decades [1–3]. The Sun, alike all other main-sequence stars, maintains a hydrostatic equilibrium balancing mechanism of the gravito-thermal type [1–4]. Incorporating the solar self-gravitational effect by applying well-known Newtonian gravity effect (encoded in the self-gravitational Poisson equation), researchers have been studying diverse solar equilibrium-fluctuation dynamical features in order to understand various solar physics problems of wide interest [1–7].

It is well-known that, in the last century, the Einsteinian General Theory of Relativity (GTR) has become well-concretized as the standard theory of gravity [8,9]. In the weak-field regime, such as the solar gravitational regime, this GTR reduces to the Newtonian gravity, which is cast in the mathematical framework of the famous Poisson equation [8]. However, the validity of this GTR-based Newtonian Poisson equation inside a self-gravitationally bound astroobject is yet to be well explored [10]. Moreover, in some regimes, Einstein's GTR theory has its own limit of validity [9–13]. Such delicate point in the Einsteinian gravitational theory lies in the gravity-matter coupling formalism [10,14,15]. In order to overcome such limitations, several alternative gravitation theories have been proposed introducing modifications in the Einstein theory [9,11,12,14–18]. It is, therefore, a matter of our great concern that extensive use of point particle-based Newtonian formalism of self-gravity is both inadequate and improper as far as seen in the literature on the grounds of modified non-local gravitational interactions [9,11,12,16,19]. Thus, in order to refine the standard Newtonian gravity formalism, different post-Newtonian approaches have already been reported in the literature [10–12,16,19–22].

This post-Newtonian approximation is applicable based on the realistic key assumption that gravitational fields inside and around the astroobject is subject to be weak and the motions of the matter within the source are very slow compared to the speed of light [16]. In this context, an attractive theory of modified gravity termed as the Eddington-inspired Born−Infeld (EiBI) theory has been put forward [10–12,19–23]. This theory is inspired by the Born–Infeld coupling action for non-linear electrodynamics and developed based on the Eddington theory of gravity [11,19,23]. The key property of the EiBI gravity theory is that it is equivalent to the GTR theory in vacuum, while, sensibly differs from the latter within the matter [10,12]. Moreover, at the classical level, this modern formulation results in a novel non-singular description of the Universe and avoids the end-stage singularity formation in a self-gravitational collapse process [10–12,24].

It is seen recently in the literature [10] that the EiBI gravity could lead to strong realistic modifications in the self-gravitationally bounded solar structure. In the classical Newtonian limit, this theory yields in a modified gravitational Poisson equation [10,12,23,24]. Thus, it could affect the evolution mechanism and the equilibrium-fluctuation features of the solar structure by enabling us to get different thermal and morphodynamical profile, and obtain modified collective acoustic mode characteristics [10,24].

After being motivated by the above realistic scenarios, we propose herein a new model approach to explore diverse solar plasma equilibrium features in light of the EiBI gravity formalism on the gravito-electrostatic sheath (GES)-based plasma framework. Unlike the previous solar GES investigations based on plasma boundary wall-interaction mechanisms amid unlike electron-ion gravito-thermal coupling interplay, as already reported in the literature [5,6,25–27], the key novelty of our work is rooted to formulate EiBI gravity-modified GES-based equilibrium solar plasma description in a thermal viscothermic model fabric. It is already reported in the literature [25,28,29] that viscosity plays a significant role in determining the transport of momentum and energy in the solar interior plasma. In our proposed study, we investigate how the inclusion of EiBI gravity and viscosity would modify the solar GES structure as well as the solar interior properties. Our speculations reveal that the EiBI gravity shifts the present solar surface boundary (SSB) outwards by 14.28% relative to the original standard Newtonian gravity-based SSB already reported in the literature [6]. Moreover, it is also seen that EiBI gravity affects diverse relevant solar parameters, such as the electrostatic potential, electric field, and Mach number defining a modified solar plasma model fabric.

Apart from the introduction part, the structural layout of this proposed manuscript is organized in a standard pattern as follows. The physical model and mathematical formalism are discussed in Section 2. The model description and governing equations are presented in Sections 2.1 and 2.2, respectively. The SSB analysis is illustrated in Section 2.3. The results and discussion are depicted in Section 3. Finally, the main conclusions drawn from our analyses are summarily presented in Section 4.

## 2. Physical model and mathematical formalism

*2.1 Model description*

The self-gravitationally bounded quasi-neutral solar interior plasma (SIP) system is considered as a weakly magnetized collisional plasma system with a spherically symmetric surface boundary of nonrigid and nonphysical nature [6,25,26]. This zeroth-order surface boundary is well-defined by an exact hydrostatic condition of gravito-electrostatic force balancing of the enclosed solar plasma mass at a numerically obtained radial position from the center of the mean solar gravitational mass [6]. For simplicity, we consider a viscous quasi-neutral unmagnetized solar interior plasma system in the spherically symmetric (reduced to radial, 1-D configuration) GES-based model comprising of mainly ionized form of hydrogen (92%) and helium (8%) [1–3,30]. The other heavy ionic and neutral species, such as α-particles, C, N, O, Fe, etc. are ignored due to their poor relative abundance (about 0.01%) [1,2,4,31]. The role of magnetic field is ignored just for mathematical simplicity in discussing the collisional solar interior plasma dynamics. The viscosity effects in our model arise due to the Coulomb collisions between the constitutive ions in the plasma, and is typically characterized by the kinematic viscosity coefficient ($\nu$), also termed as the momentum diffusivity [25,29]. It fairly apes the realistic solar plasmas and circumvent physical environments as extensively found in the literature [29-35].

In order to execute our scale-free calculation ansatz, various standard notations and symbols relevant for describing the entire solar plasma system are given in Appendix A [26,27,32]. Applying this customary symbolism from Appendix A, quite pertinent to the helio-plasma dynamics [6,26,27,33], we describe our espoused standard astrophysical normalization scheme with all the usual Jeansean notations and significances as in Appendix B. The solar interior plasma system, confined hydrodynamically under the action of self-gravity in a spherically symmetric geometry, justifiably enables us to employ the Jeansean spatial scales for the normalization scheme. The applicability of the global quasi-neutral nature ($n_e \approx n_i = n$) of the entire solar plasma system is admissibly based on the representative ground that the asymptotic value of the spatial Debye-to-Jeans length scale ratio [6,25] is almost zero $\left(\lambda_{D_e}/\lambda_J \approx 10^{-20} \sim 0\right)$.

## 2.2 Governing equations

We consider that our proposed solar interior plasma volume is composed of the thermal (Maxwellian) electrons and inertial ions [6]. Thus, the usual unnormalized form of the Maxwell-Boltzmann density distribution describing plasma thermal electrons in the solar plasma is given with all the customary notations as

$$n_e = n_0 e^{\frac{e\phi}{T_e}} \qquad (1)$$

Here, $n_e$ denotes the unnormalized electron number density, and $n_0 = \rho_\odot/m_i$ defines the average bulk density of the solar interior plasma.

The ions in our model are described by their full inertial response dynamics. The equations governing the ion flow dynamics are thus include the ion momentum equation as well as the ion continuity equations. Thus, the unnormalized forms of these equations with all the usual notations are respectively cast as

$$m_i n_i \left[\frac{\partial \vec{v}_i}{\partial t} + (\vec{v}_i \cdot \vec{\nabla})\vec{v}_i\right] = q_i n_i \vec{E} - \vec{\nabla} p_i + m_i n_i \vec{g} + m_i n_i \nu \nabla^2 \vec{v}_i, \qquad (2)$$

$$\frac{\partial n_i}{\partial t} + \vec{\nabla} \cdot (n_i \vec{v}_i) = 0. \qquad (3)$$

Here, $\vec{E}$, $p_i$, $\vec{g}$, and $\vec{v}_i$ stand for the effective electric field, thermal pressure, gravitational field, and the ionic velocity, respectively. The term $\nu$ denotes the zeroth-order kinematic viscosity coefficient and is given with all usual notations as $\nu = 0.96(n_i k_B T_i \tau_i/\rho_\odot)$ m$^2$ s$^{-1}$, where, $\tau_i = 0.75(T_i^{3/2} n_i^{-1})$ s represents the typical Coulombic relaxation time for the ions [25,28,29]. Hence, one can obtain $\nu = 0.72(k_B T_i^{5/2}/\rho_\odot)$ m$^2$ s$^{-1}$ (all in SI units).

The constitutive species (thermal electrons and inertial ions) are coupled by the gravito-electrostatic Poisson formalism which complements the steady state hydrodynamic configurations giving a complete description of the GES structure around the SSB [6]. Thus, the unnormalized electrostatic Poisson equation with all the usual notations cast as

$$\nabla^2 \phi = -\frac{\rho_q}{\epsilon_0} \qquad (4)$$

The EiBI-modified Poisson equation governing the self-gravitationally bounded solar interior plasma system in a weak-field, slow-motion approximation limit is given with all the customary notations as

$$\nabla^2 \psi = 4\pi G \rho_m + \frac{\chi}{4} \nabla^2 \rho_m \qquad (5)$$

Here, the term $4\pi G \rho_m$ on the right-hand side stands for the standard Newtonian contribution, while the second term i.e., $\frac{\chi}{4}\nabla^2 \rho_m$ originates due to the correction brought by the EiBI gravity effect in the non-relativistic Newtonian limit defining solar plasma system, where the term $\chi$ is coined as the EiBI gravity parameter [10,19–23]. It is reported in the recent literature that an astroobject with radius $R$ holding its mass together by its own self-gravity, leads to the constraint $\chi \leq GR^2$ [13]. Hence, for the solar interior plasma mass distribution, one can easily estimate this value to be $\chi \leq 10^8$ kg$^{-1}$ m$^5$ s$^{-2}$ [13].

Applying the relevant astrophysical normalization scheme as mentioned in Appendix B, the above set of SIP governing Eqs. (1)-(5) in the time-independent normalized form with all the usual notations and symbolism are systematically reproduced and recast, respectively, as

$$N_e = e^{\Phi},\tag{6}$$

$$M\frac{dM}{d\xi} = -\frac{d\Phi}{d\xi} - \epsilon_T \frac{1}{N_i}\frac{dN_i}{d\xi} - \frac{d\Psi}{d\xi} + \frac{\nu}{c_S \lambda_J}\left[\frac{d^2 M}{d\xi^2} + \frac{2}{\xi}\frac{dM}{d\xi} - \frac{2M}{\xi^2}\right],\tag{7}$$

$$\frac{1}{N_i}\frac{dN_i}{d\xi} + \frac{1}{M}\frac{dM}{d\xi} + \frac{2}{\xi} = 0,\tag{8}$$

$$\frac{d^2\Psi}{d\xi^2} + \frac{2}{\xi}\frac{d\Psi}{d\xi} = N_i + \frac{1}{16\pi G}\frac{1}{\lambda_J^2}\chi\left[\frac{d^2 N_i}{d\xi^2} + \frac{2}{\xi}\frac{dN_i}{d\xi}\right],\tag{9}$$

$$\left(\frac{\lambda_{D_e}}{\lambda_J}\right)^2\left[\frac{d^2\Phi}{d\xi^2} + \frac{2}{\xi}\frac{d\Phi}{d\xi}\right] = N_e - N_i.\tag{10}$$

Here, the minus sign appears in the gravitational term $d\Psi/d\xi$ to indicate the radially inward solar self-gravity. Besides, $\epsilon_T = T_i/T_e$ represents the solar ion-to-electron temperature ratio and $\lambda_{D_e} = \left(\frac{\epsilon_0 T_e}{n_0 e^2}\right)^{1/2}$ denotes the plasma electron Debye length of the defined solar interior plasma system.

Thus, Eqs. (6)-(10) comprise a completely closed set of normalized basic governing equations which define the basic physics of the GES-based bounded solar interior plasma features modified with the EiBI gravity effect. As already mentioned earlier, on the typical gravitational scale length of the solar plasma system, one can obtain the realistic (physical) approximation $\lambda_{D_e}/\lambda_J \approx 10^{-20} \sim 0$ implying that the plasma electron Debye length $(\lambda_{D_e})$ is much smaller than the Jeans scale length $(\lambda_J)$. This reduces Eq. (10) into the well-known quasi-neutral condition cast as

$$N_e = N_i = N = e^{\Phi}\tag{11}$$

However, it does not indicate that the ions are also governed by the Boltzmannian density distribution. Equation (11) can be further rewritten in a differential form as

$$\frac{1}{N}\frac{dN}{d\xi} = \frac{d\Phi}{d\xi}\tag{12}$$

Using Eqs. (6)-(12) we finally obtain the reduced form of the basic set of coupled non-linear SIP governing equations with all the usual notations given as

$$(M^2 - \epsilon_T)\frac{1}{M}\frac{dM}{d\xi} = -\frac{d\Phi}{d\xi} + \epsilon_T \frac{2}{\xi} - g_s + \frac{\nu}{c_S \lambda_J}\left[\frac{d^2 M}{d\xi^2} + \frac{2}{\xi}\frac{dM}{d\xi} - \frac{2M}{\xi^2}\right],\tag{13}$$

$$\frac{d\Phi}{d\xi} + \frac{1}{M}\frac{dM}{d\xi} + \frac{2}{\xi} = 0,\tag{14}$$

$$\frac{dg_s}{d\xi} + \frac{2}{\xi}g_s = e^{\Phi} + \frac{1}{16\pi G}\frac{1}{\lambda_J^2}\chi\left[e^{\Phi}\left\{\frac{d^2\Phi}{d\xi^2} + \left(\frac{d\Phi}{d\xi}\right)^2\right\} + \frac{2}{\xi}e^{\Phi}\frac{d\Phi}{d\xi}\right].\tag{15}$$

Here, for the sake of the simplicity of analysis, we denote the solar self-gravitational acceleration as $g_s = d\Psi/d\xi$.

This set of coupled non-linear differential equations constituting a closed dynamical system of governing equations can be now used to analyse the EiBI gravity-modified GES structure features of the bounded solar interior plasma system.

*2.3 SSB analysis*

### 2.3.1 Numerical analysis

In order to solve the set of coupled non-linear dynamical evolution equations (Eqs. (13)-(15)), we have to determine an autonomous set of initial values of the relevant defined solar physical variables. For the initial-value problem at hand, we need to derive the prerequisite initial values of the physical variables $M(\xi)$, $g_s(\xi)$, and $\Phi(\xi)$ inside the solar interior. We obtain the self-consistent choice of the initial values by putting $dg_s/d\xi|_{\xi_i} = 0$, $d\Phi/d\xi|_{\xi_i} = 0$, and $d^2\Phi/d\xi^2|_{\xi_i} = 0$ at Eqs. (13)-(15). Hence, the expressions for a physically valid set of initial values of the given physical variables are obtained with all the customary notations as

$$g_i = \frac{1}{2}\xi_i e^{\Phi_i}, \tag{16}$$

$$\left.\frac{dM}{d\xi}\right|_{\xi_i} = -e^{\frac{\Phi_i}{2}}, \tag{17}$$

$$M_i = \frac{1}{2}\xi_i e^{\frac{\Phi_i}{2}}, \tag{18}$$

It is noteworthy that the values of $\xi_i$ and $\Phi_i$ are chosen arbitrarily. The initial values of all the other relevant solar plasma variables are systematically calculated by using the set of Eqs. (16)-(18) and summarily are shown in Table 1 as follows.

**Table 1: Initial and boundary values of physical parameters**

| Parameter | At initial point | At SSB | Initial value |
|---|---|---|---|
| Radial distance ($\xi$) | $\xi_i = 0$ | $\xi_{SSB} = 4$ | $\xi_i$, arbitrary |
| Electrostatic potential ($\Phi$) | $d\Phi/d\xi|_{\xi_i} = 0$ | $\Phi_{SSB} \sim -1.17$, $d\Phi/d\xi|_{SSB} \sim -0.60$ | $\Phi_i$, arbitrary |
| Self-gravity ($g_s$) | $dg_s/d\xi|_{\xi_i} = 0$ | $g_{SSB} \sim 0.61$, $dg_s/d\xi|_{SSB} = 0$ | $g_i = \frac{1}{2}\xi_i e^{\Phi_i}$, derived |
| Mach number ($M$) | $dM/d\xi|_{\xi_i} = -e^{\Phi_i/2}$ | $M_{SSB} = 10^{-6}$, $dM/d\xi|_{SSB} = 0$ | $M_i = \frac{1}{2}\xi_i e^{\frac{\Phi_i}{2}}$, derived |

Now, we apply the well-known fourth-order Runge−Kutta (RK-IV) method upon the Eqs. (13)-(15) to study the diversified features of modified SSB. It is interestingly found that the EiBI gravity-modified solar radius $(R_{SSB_M})$ is equal to 4 times of the Jeans length $(\lambda_J)$ for the mean solar mass density $(\rho_\odot = 1.43 \times 10^3$ kg m$^{-3})$. Hence, we can rewrite this in a mathematical form as $R_{SSB_M} = 4\lambda_J = 12.4 \times 10^8$ m. A schematic diagram has been portrayed to show the comparative structure of the original Newtonian gravity-based and EiBI gravity modified solar GES model in Fig. 1.

Performing constructive numerical analysis, we also find that the value of electrostatic potential at SSB is about $\Phi \sim -1.17$, which is about $-1.17$ kV. The SSB also acquires the maximum value of the solar interior plasma mass self-gravity, $g_{SSB} \sim 0.61$. The electrostatic potential gradient at the surface is found to be $d\Phi/d\xi|_{SSB} = E_{SSB} \sim -0.6$. It indicates that that the strengths of the GES-associated self-gravity and electric field are almost equal satisfying the gravito-electrostatic force balancing mechanism justifiably on the solar interior plasma system.

### 2.3.2 Properties of modified SSB

A constructive numerical analysis enables us to obtain the boundary values of the EiBI gravity-modified relevant solar parameter defining the SSB as listed in Table 2. We portray different numerical plots (Figs. 2-6) to depict the modified SSB description. It is numerically obtained from the plot that at the specifically defined modified SSB, the Mach number comes out to be minimum with a non-zero value of the order of $10^{-6}$ (Fig. 6). Using this value, at SSB, Eq. (13) can be reduced to $dM/d\xi \approx -M/\xi_{SSB} = -2.5 \times 10^{-7} \sim 0$. It is associated to a quasi-hydrostatic type of equilibrium condition arises due to the gravito-electrostatic force balancing mechanism on the SSB with outward solar interior plasma flow having minimum speed of about 0.3 m s$^{-1}$. Hence, one can demonstrate the fact that the interaction of the solar interior plasma gives rise to the solar wind plasma flow.

The time-independent solar self-gravity $g_s$ associated with the solar interior plasma mass distribution is now modified by EiBI gravity and is found to have its maximum potential at EiBI gravity-modified SSB which now lies at a radial distance of $\xi = \xi_{SSB} = 4$ (Fig. 2). One can note that the quasi-neutral hydrostatic equilibrium condition is held quite good at this boundary with the relation $g_{SSB} \approx |d\Phi/d\xi|_{SSB}|$, as depicted in Fig. 2.

The EiBI modified GES-based radial size ($1R_\odot$) of the spherically symmetric solar plasma chamber is obtained to be $R_{SSB_M} = 4\lambda_J = 12.4 \times 10^8$ m (Fig. 1). Hence, it is obvious that the EiBI modified solar radius is increased by 14.28% relative to the original Newtonian gravity-based SSB, $R_{SSB} = 3.5\lambda_J$. Also, one can easily find that it is larger than the standard solar radius, $R_{SSM} = 6.96 \times 10^8$ m by about 78%.

## 3. Results and discussions

The effect of EiBI gravity on the GES structure features of the viscous solar interior plasma is numerically analysed. The basic equations governing the SIP in normalized form are formulated. Employing RK-IV method on the governing equations and performing some numerical analysis, we depict different colourspectral profiles (Figs. 2-6) in order to show the effect of the EiBI gravity parameter on the GES-based solar interior plasma features. Our theoretical predictions redefine the GES structure features modified with EiBI gravity. The quasi-hydrostatic equilibrium condition characterizes and ensures the formation of the modified GES-structure as well as the SSB. Keeping the solar ion-to-electron temperature fixed at $\epsilon_T = 0.4$, we numerically obtain the values of different relevant parameters at SSB and summarily portray in Table 1. Such numerical values, $\Phi_{SSB} \sim -1.17$, $g_{SSB} \sim 0.61$, $d\Phi/d\xi|_{SSB} \sim -0.6$, $M_{SSB} \sim 10^{-6}$ define the characteristics of our proposed EiBI gravity-modified solar surface boundary.

Figure 2 depicts that the quasi-hydrostatic condition is well satisfied with the relation $g_{SSB} \approx |d\Phi/d\xi|_{SSB}| \sim 0.6$ at the EiBI gravity-modified different SSB lies at $\xi = 4$. Hence, in our model, the inclusion of EiBI gravity and viscous effect shifts the SSB outwards by 14.28% relative to the original Newtonian gravity-based SSB as already reported in the literature [6]. We can speculate that this shifting occurs due to the consideration of modified gravity-matter coupling effect brought by the $\chi$ parameter in our GES model scenario.

Figure 3 shows the profile of $g_s$ associated with the SIP flow dynamics with variation in $\xi$ for different judicious indicated values of the EiBI gravity parameter ($\chi$) as indicated. It is interestingly seen that the radial position of the EiBI gravity-modified SSB from the heliocentre ($\xi = 0$) is increasing as the $\chi$ increasing. It explicitly indicates the fact that as $\chi$ increases, the SSB shifts outwards in order to maintain the hydrostatic equilibrium condition.

In Figure 4, we depict the profile of $g_s$ associated with the SIP flow dynamics with variation in $\xi$-scale for different indicated values of the $\chi$-parameter. It is seen that, as the $\chi$-parameter increases, the normalized $E$-field magnitude $|d\Phi/d\xi|$ shows a slight decreasing tendency.

Figure 5 portrays the profile of the normalized electrostatic potential ($\Phi$) associated with the SIP flow dynamics with variation in $\xi$ for different judicious indicated values of the $\chi$-parameter. It is observed that $\Phi$ shows very slight decreasing propensity as the $\chi$-value increases.

In Figure 6, we exhibit the profile of $M$ associated with the SIP flow dynamics with variation in the $\xi$-scale for different values of the $\chi$-parameter as indicated in the figure. It is seen interestingly that $M$ has almost no significant variation with $\chi$-values.

In our proposed model, we study the effect of the EiBI gravity on different relevant parameters of the solar interior plasma on the GES framework. The main comparison between the previously reported Newtonian gravity-based GES and our EiBI gravity-based one is summarily given in Table 2.

Table 2: Parametric contrast between the Newtonian and EiBI gravity-based GES analyses

| Parameters at SSB | Newtonian GES | EiBI GES | Remark |
|---|---|---|---|
| Normalized position ($\xi_{SSB}$) | 3.50 | 4.00 | Increased by 14.28% |
| Unnormalized position ($r_{SSB}$) | $10.8 \times 10^8$ m | $12.4 \times 10^8$ m | |
| Normalized potential ($\Phi_{SSB}$) | −1.00 | −1.17 | Increased by 17% |
| Unnormalized potential ($\phi_{SSB}$) | −100 V | −117 V | |
| Normalized self-gravity ($g_{SSB}$) | 0.60 | 0.61 | Increased by 1.78% |
| Unnormalized self-gravity ($g'_{SSB}$) | 185.8 m s$^{-2}$ | 189.1 m s$^{-2}$ | |
| Normalized $E$-field ($d\Phi/d\xi|_{SSB}$) | −0.62 | −0.60 | Decreased by 3.23% |
| Unnormalized $E$-field ($d\phi/dr|_{SSB}$) | $-2 \times 10^{-7}$ N C$^{-1}$ | $-1.9 \times 10^{-7}$ N C$^{-1}$ | |
| Mach number ($M_{SSB}$) | $10^{-7}$ | $10^{-6}$ | Increased by 900% |
| Unnormalized velocity ($v_{SSB}$) | $3 \times 10^{-2}$ m s$^{-1}$ | $3 \times 10^{-1}$ m s$^{-1}$ | |

## 4. Conclusions

In this methodical study, we report a detailed time-independent (equilibrium) analysis of the EiBI gravity-modified self-gravitationally bounded solar interior plasma features founded on the GES model framework on the relevant solar (Jeansean) spatial scales. It incorporates the conjoint effects of the EiBI gravity and viscothermic action for the first time to study the solar plasma equilibrium features in the GES model framework. The EiBI gravity parameter, $\chi$, in our solar plasma model brings about the real-time modification in the gravity-matter coupling formalism. A constructive mathematical analysis is carried out over the helio-structure governing equations of the self-gravitationally confined SIP system. We systematically apply the RK-IV method in order to obtain the exact solution of the basic set of the governing equations. Performing a detailed numerical analysis, we reveal different colourspectral profiles (Figs. 2-6) depicting the EiBI gravity-modified equilibrium GES features. It is interestingly found that the EiBI gravity-modified SSB is shifted outside by 14.28% relative to the original Newtonian gravity-based one. This shifting occurs due to the realistic modification in the solar interior gravity-matter coupling formalism yielded by the EiBI gravity.

It may be noteworthy that, for $\chi \gtrsim 0.02 GR_\odot^2$, the solar (SIP) hydrodynamic equilibrium configuration remains well set up. However, its internal gravity-matter coupling formalism becomes so sensibly modified due to this $\chi$-value that the estimated relevant solar parameters, such as the solar surface heavy-element content, age, radius, mass, and luminosity may deviate from the standard solar model predictions [10,34]. As a consequence, when the other solar parameters are kept fixed in investigating the $\chi$-dependent SSB features, the SSB gets shifted outwards to maintain the required equilibrium configuration in order to form the EiBI-gravity modified GES structure for smooth and monotonous plasma parametric variations in accordance with the basic energy-flux conservation laws. Besides, the physical value of the EiBI gravity-modified self-gravity at SSB is found to be $g'_{SSB} = 189.1$ m s$^{-2}$, which is now increased by 1.78% relative to the Newtonian one, showing almost a fair corroboration with the standard solar surface self-gravity value, $g_{SSM} = 274$ m s$^{-2}$ [1,2]. In addition, our predicted solar surface $E$-field value is quite in a good agreement with the previously predicted $E$-field data as reported in the literature [1,2].

One can also note how the inclusion of the EiBI gravity and viscosity affects the other relevant solar parameters as summarized in Table 2 with comparing it with the original Newtonian gravity-based one. Table 2 gives a synoptic view of both the normalized and unnormalized physical values of the relevant solar plasma parameters at the Newtonian- and EiBI-based SSBs along with the corresponding changes. It is observed that at the diffused SSB, formed under the GES formation framework, the solar self-gravity potential becomes maximum and is balanced with the solar interior $E$-field strength, i.e., $g_{SSB} \approx |d\Phi/d\xi| \sim 0.6$. It is pertinent to add here that at this EiBI gravity-modified SSB, an outward solar interior plasma outflow occurs with a reduced minimum speed of $M_{SSB} = 10^{-6}$ characterized by the quasi-hydrostatic equilibrium condition. It concretizes the interesting fact that the genesis of the subsonic solar wind plasma at the SSB lies actually in the SIP dynamics as seen in the literature [6,25,27,33]. Our analyses may interestingly provide new physical insights for understanding the realistic equilibrium dynamics of the GES-based SIP structure in the light of the EiBI gravity-modified gravity-matter coupling formalism, which, in fact, has never been well addressed before in the earlier solar studies.

It is worth mentioning here that, as the SIP self-gravity (internal) transforms spatially into the SWP gravity (external), the role of the modified gravitational Poisson equation in the subsequent SWP study becomes redundant. Applying different forms of the non-Maxwellian distributions (non-thermal), we could further explore the non-thermal features of the solar plasma dynamics in such circumstances [35,36]. In addition, considering both the constitutive Helium and Hydrogen ion fields, this model formalism can be further refined to understand diversified solar electrostatic instabilities in a realistic integrated manner [30].

The Plasma Analyzer Package for Aditya (PAPA) payload on the ongoing Aditya-L1 mission by the Indian Space Research Organization (ISRO) aims at manifold in-situ observations of the solar wind composition, associated various collective waves, and wind particle energy distribution [37]. Our proposed theoretical analyses could thereby be helpful in bridging tentative correlation and consistency of the predicted GES-based solar plasma features with both existing and forthcoming solar observational data [1,37,38]. It is believed that this semi-analytic investigation could fairly strengthen our conceptional framework on the entire solar plasma dynamics and structural evolution. It is finally anticipated that our model analyses could justifiably have broader applicability to study the solar morphodynamical structure, its circumambient atmosphere, and its evolution mechanism with a proper incorporation of the post-Newtonian gravity effects founded on a more realistic astronomical fabric.

**Declaration of competing interest**

The authors declare that there is no known competing financial interests or personal relationships that could have appeared to influence the work reported in this paper.

**Data availability statement**

Data could be made available on reasonable request to the corresponding author.

**Acknowledgements**

The authors duly acknowledge the active cooperation availed from Tezpur University. The dynamic support from the colleagues of the Astrophysical Plasma and Nonlinear Dynamics Research Laboratory (APNDRL), Department of Physics, Tezpur University, is worth mentioning. The support received (by PKK) through the SERB Project, Government of India, is duly recognized. The award of IUCAA-Associateship (Inter-University Centre for Astronomy and Astrophysics, Pune, India, to PKK) is gratefully appreciated. SD would like to thankfully acknowledge the receipt of DST-INSPIRE research fellowship (Grant: DST/INSPIRE Fellowship/2021/IF210234).

# Appendices

## Appendix A. ADOPTED RELEVANT NOTATIONS AND SYMBOLS

| Physical parameter | Symbol | Magnitude |
|---|---|---|
| Electron mass | $m_e$ | $9.31 \times 10^{-31}$ kg |
| Ion mass | $m_i$ | $1.67 \times 10^{-27}$ kg |
| Electron charge | $q_e = -e$ | $-1.6 \times 10^{-19}$ C |
| Ion charge | $q_i = +e$ | $+1.6 \times 10^{-19}$ C |
| Electron temperature | $T_e$ | $10^2$ eV |
| Ion temperature | $T_i$ | 10 eV |
| Jeans frequency | $\omega_J = c_s/\lambda_J$ | $10^{-3}$ s |
| Universal gravitational constant | $G$ | $6.67 \times 10^{-11}$ m³ kg⁻¹ s⁻² |
| Mean solar mass | $M_\odot$ | $2 \times 10^{30}$ kg |
| Mean solar plasma mass density | $\rho_\odot$ | $1.43 \times 10^3$ kg m⁻³ |

## Appendix B. ADOPTED ASTROPHYSICAL NORMALIZATION SCHEME

| Normalized parameter | Normalizing parameter | Magnitude |
|---|---|---|
| Radial distance $(\xi = r/\lambda_J)$ | Jeans length $(\lambda_J)$ | $3.1 \times 10^8$ m |
| Population density $(N_{e(i)} = n_{e(i)}/n_0)$ | Mean SIP number density $(n_0)$ | $10^{30}$ m⁻³ |
| Mach number $(M_i = v_i/c_s)$ | Sound phase speed $(c_s)$ | $3.1 \times 10^5$ m s⁻¹ |
| Gravitational potential $(\Psi = \psi/c_s^2)$ | Sound phase speed squared $(c_s^2)$ | $9.5 \times 10^{10}$ m² s⁻² |
| Electrostatic potential $(\Phi = e\phi/T_e)$ | Thermal potential $(T_e/e)$ | $10^2$ J C⁻¹ |

**Figures**

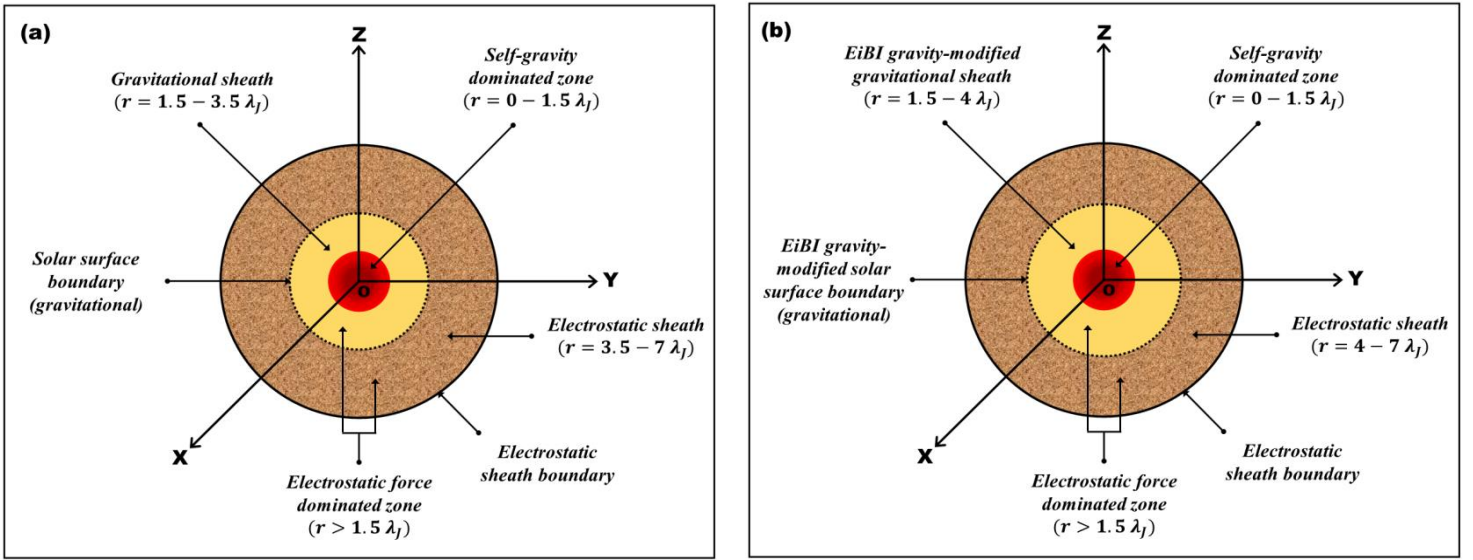

**Fig. 1.** Schematic diagram of the Sun and its circumambient atmosphere according to the (a) Newtonian gravity-based GES model and (b) EiBI gravity-based GES model. Different concentric constitutive layers of the models are depicted separately in a similar footing.

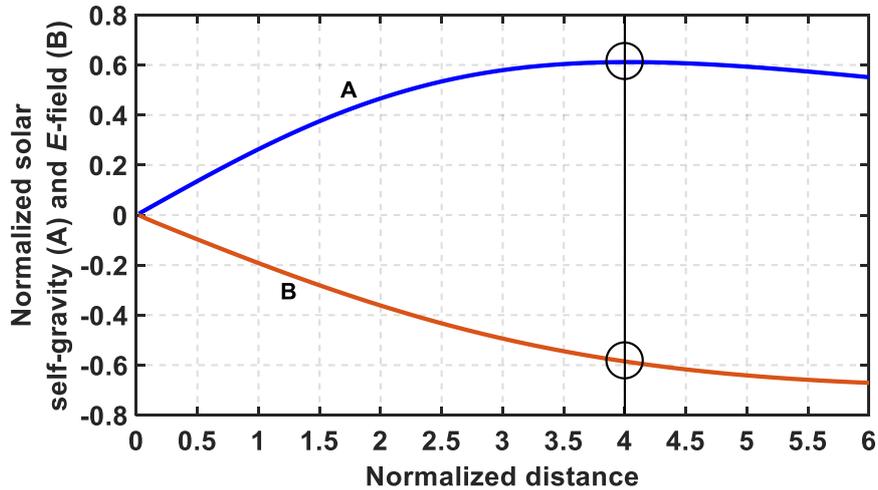

**Fig. 2.** Profile of the normalized solar (A) Self-gravity ($g$) and (B) Electric field strength ($d\Phi/d\xi$) associated with the SIP flow dynamics with variation in the Jeans-normalized radial distance ($\xi$) from the heliocenter ($\xi = 0$), specifying the exact EiBI gravity-modified SSB location at $\xi = 4$, with different judicious input and initial values presented in the text.

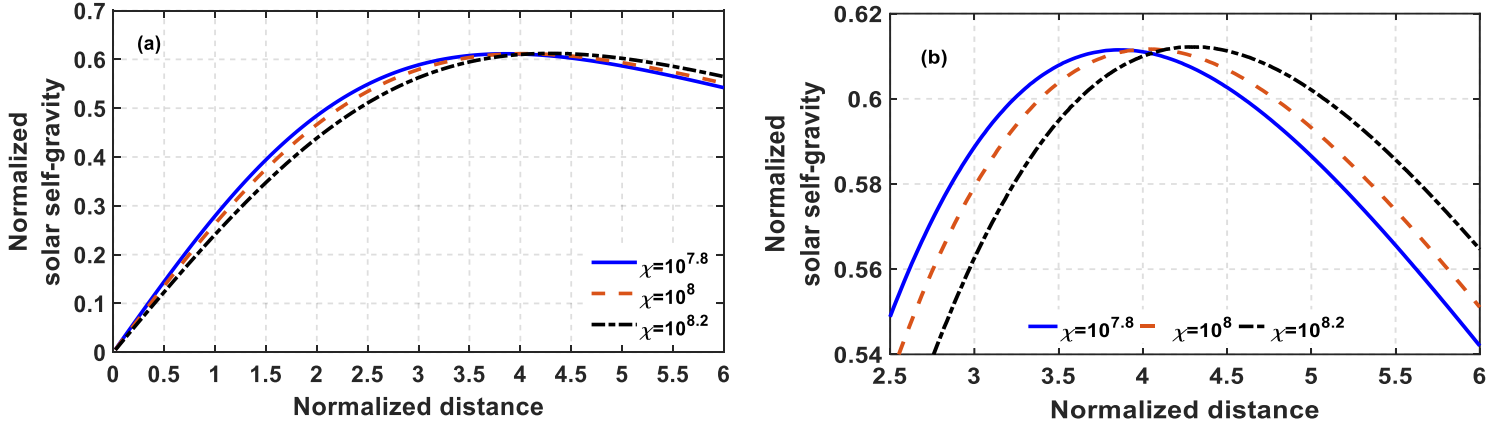

**Fig. 3.** Profile of the normalized (a) Solar self-gravity ($g$) associated with the SIP flow dynamics with variation in the Jeans-normalized radial distance ($\xi$) from the heliocenter ($\xi = 0$) for different judicious indicated values of the EiBI gravity parameter ($\chi$) and (b) Zoomed-in part of the same detailing the SSB features.

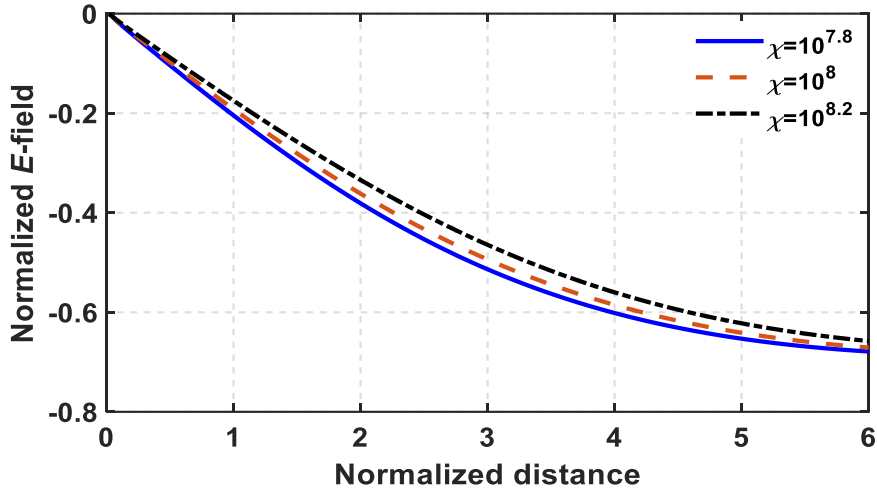

**Fig. 4.** Profile of the normalized electric field strength ($d\Phi/d\xi$) associated with the SIP flow dynamics with variation in the Jeans-normalized radial distance ($\xi$) from the heliocenter ($\xi = 0$) for different judicious indicated values of the EiBI gravity parameter ($\chi$).

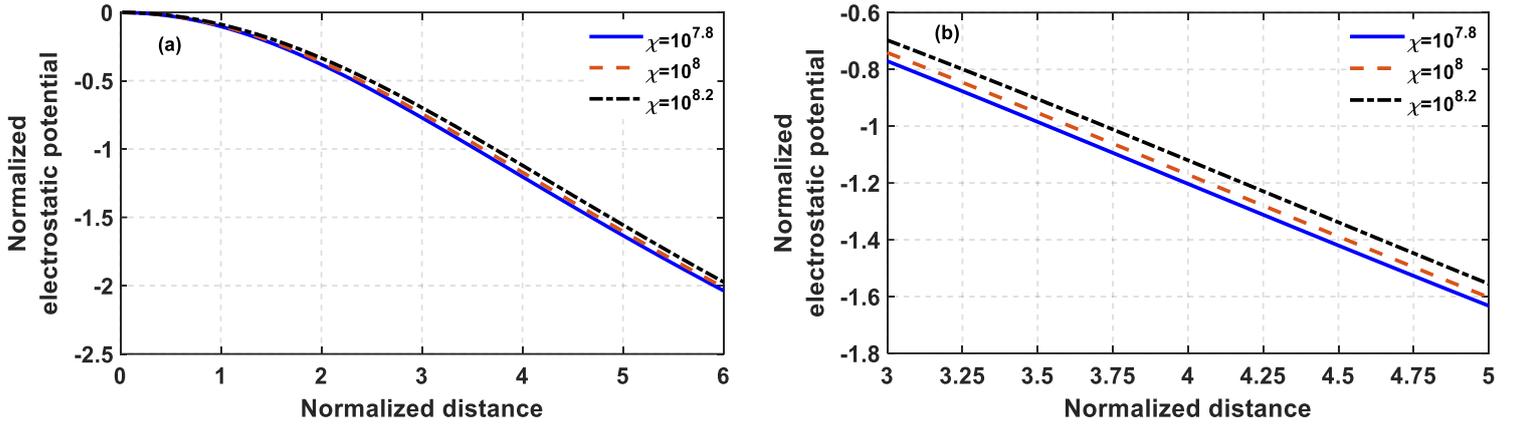

**Fig. 5.** Profile of the normalized electrostatic potential ($\Phi$) associated with the SIP flow dynamics with variation in the Jeans-normalized radial distance ($\xi$) from the heliocenter ($\xi = 0$) for different judicious indicated values of the EiBI gravity parameter ($\chi$) and (b) Zoomed-in part of the same detailing the SSB features.

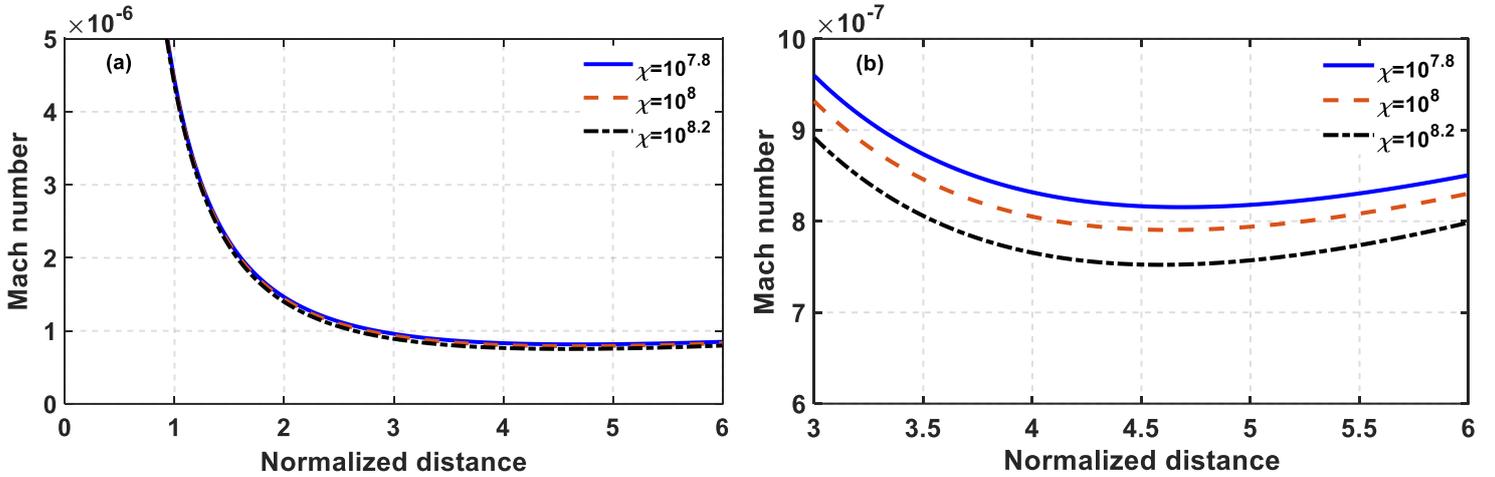

**Fig. 6.** Profile of the normalized (a) Mach number ($M$) associated with the SIP flow dynamics with variation in the Jeans-normalized radial distance ($\xi$) from the heliocenter ($\xi = 0$) for different indicated judicious values of the EiBI gravity parameter ($\chi$) and (b) Zoomed-in part of the same detailing the SSB features.